# Multimode optically adiabatic silica glass submicron taper

Chang Kyun Ha, Kee Hwan Nam, and Myeong Soo Kang[*]

*Department of Physics, Korea Advanced Institute of Science and Technology (KAIST)*

*291 Daehak-ro, Yuseong-gu, Daejeon 34141, Republic of Korea*

[*]*mskang@kaist.ac.kr*

Silica nanofibers fabricated by tapering optical fibers have attracted considerable interest as an ultimate platform for high-efficiency light-matter interactions. While previously demonstrated applications relied exclusively on the low-loss transmission of only the fundamental mode, the implementation of multimode tapers that adiabatically transmit several modes has remained very challenging, hindering their use in various emerging applications in multimode nonlinear optics and quantum optics. Here, we report the first realization of multimode submicron tapers that permit the simultaneous adiabatic transmission of multiple higher-order modes including the LP$_{02}$ mode, through introducing deep wet-etching of conventional fiber before fiber tapering. Furthermore, as a critical application, we demonstrate "fundamental-to-fundamental" all-fiber third-harmonic generation with high conversion efficiencies. Our work paves the way for ultrahigh-efficiency multimode nonlinear and quantum optics, facilitating nonclassical light generation in the multimode regime, multimode soliton interactions and photonic quantum gates, and manipulation of the evanescent-field-induced optical trapping potentials of atoms and nanoparticles.

# INTRODUCTION

Optical micro/nano-tapers (OMNTs), which can guide light along the subwavelength-diameter waist [1,2], have been utilized as a versatile photonic platform in a vast range of applications. The significant evanescent field can be exploited to achieve near-unity-efficiency evanescent light coupling to other photonic systems [3–5], tight optical trapping of atoms [6–8] and nanoparticles [9,10] in the vicinity of the waist for their efficient interactions with the guided light, as well as high-sensitivity optical detection of gases and chemicals [11,12]. Various nonlinear and quantum optical phenomena in OMNTs have also attracted growing interest. OMNTs provide not only tight confinement of light within a tiny cross-section, which yields radically enhanced effective optical nonlinearities, but also a unique capability of simple and broad dispersion engineering through varying the waist thickness. These properties have enabled broader-than-octave-spanning supercontinuum generation [13], optical excitation of strongly confined coherent acoustic phonons [14], and creation of correlated photon pairs via spontaneous four-wave mixing [15,16].

While these experiments were demonstrated using the light guided in the fundamental $LP_{01}$ mode only, several emerging applications can be facilitated exclusively by utilizing the higher-order modes (HOMs), to access substantially broad landscapes of optical dispersion for intermodal phase-matching of multimode nonlinear and quantum optical processes and to engineer the evanescent field patterns for controlling the spatial profiles of optical trapping potentials. For instance, harmonic generation [17–21] and parametric down-conversion [22,23] in OMNTs demand phase-matching between the fundamental wave in the $LP_{01}$ mode and the harmonic in the HOM [24]. Photon-pair generation via intermodal four-wave mixing [25] and multimode soliton dynamics such as single-pump optical event horizon interactions [26] and soliton self-mode conversion [27] have also been proposed as novel phenomena involving the HOMs.

All these applications require low-loss robust transmission of multiple HOMs, particularly the $LP_{0m}$ modes ($m>1$) having the same azimuthal dependence of the field distribution as the $LP_{01}$ mode, in OMNTs. However, the implementation of such multimode OMNTs has remained very challenging. Some recent experimental demonstrations have been limited to two-mode OMNTs that adiabatically transmit the $LP_{01}$ and the $LP_{11}$ modes only, relying on the use of specially designed fiber with a reduced cladding [28,29] or an unconventional index distribution [30,31]. In this paper, we report the first realization of multimode adiabatic submicron tapers (MASTs) that permit the low-loss adiabatic transmission of the $LP_{02}$ mode together with the $LP_{01}$, $LP_{11}$, and $LP_{21}$ modes. In contrast to the popular method of direct tapering of optical fiber, we employ a two-step process: deep wet-etching optical fiber to a carefully predetermined cladding diameter below ~20 μm and tapering the etched fiber down to the targeted submicron waist diameter. Our fabrication method works even with commercial off-the-shelf telecom step-index fiber, offering excellent compatibility with standard fiber-optic systems. Furthermore, as an immediate and critical application of our MASTs, we demonstrate the first 'fundamental-to-fundamental' all-fiber phase-matched third-harmonic generation (THG) with high conversion efficiencies ($>10^{-4}$). This process is the very prerequisite to realizing the exact reverse, the 'fundamental-to-fundamental' third-order parametric down-conversion [22,23], which has recently received considerable attention as a new type of nonclassical light source, although this is yet to be experimentally demonstrated.

## RESULTS

**Multimode adiabatic submicron taper (MAST)**

Figure 1 describes the operating principle and numerical design of a silica MAST. Silica OMNTs can be fabricated by heating and pulling a section of unjacketed silica optical fiber [32], which creates a pair of transition regions connecting the taper waist and intact optical fiber. For the adiabatic guidance of an optical mode along the OMNT, the effective index of the mode should be kept sufficiently distinct from those of the other modes in the transition region [32]. However, an ultimate limitation in the implementation of MASTs with tapering conventional optical fiber is that HOMs become very close to each other in terms of effective index right after they are transformed into the cladding modes at the taper transition. Figure 1(a) shows the calculated effective indices of several optical modes at 532 nm wavelength over a range of taper diameter in the case when a section of 125-μm-thick optical fiber is tapered. The $HE_{11}$ ($LP_{01}$) mode is well isolated from all HOMs. As a result, it transmits adiabatically through the taper at any diameter for ordinary (i.e., not too steep) taper transitions [32]. In sharp contrast, the effective indices of the HOMs get very similar to each other when they are guided as the cladding modes near the core-mode cut-off diameter. The inset in Fig. 1(a) highlights the case of the three hybrid HOMs of our primary interest in this work – the $EH_{11}$ and the $HE_{31}$ modes (belonging to $LP_{21}$), and the $HE_{12}$ ($LP_{02}$) mode. The $HE_{12}$ mode becomes almost degenerate with both the $EH_{11}$ and the $HE_{31}$ modes around the taper diameter of 71.6 μm. It should be noted that the $EH_{11}$ and the $HE_{12}$ modes exhibit anti-crossing, their intermodal beatlength markedly larger than 1 m as shown in Fig. 1(b). Furthermore, these modes closely approach the $EH_{12}$ and the $HE_{13}$ modes that are already cladding-guided in the intact (non-tapered) fiber. The $EH_{11}$ and the $HE_{12}$ modes display another anti-crossing at the taper thickness of 11.8 μm, where the core becomes so thin that its role in the light guidance becomes negligible. Such complicated structures of modal anti-crossing, as well as crossing, emerge for many pairs of hybrid HOMs, in particular for the $EH_{l,m}$ (belonging to $LP_{l+1,m}$) and the $HE_{l,m+1}$ (belonging to $LP_{l-1,m+1}$) modes ($l$, $m$: positive integers), at specific taper diameters, which hinders their low-loss adiabatic transmission.

We overcome this bottleneck by significantly reducing the initial cladding diameter before the tapering process. Figure 1(c) displays the calculated effective indices in the scenario when the cladding diameter is decreased to 17 μm prior to tapering, while leaving the initial core parameters unchanged. The near-degeneracy among the cladding-guided HOMs is strongly lifted, and as a result, each mode is well separated from other modes at any taper diameter. In particular, the anti-crossing of the $EH_{11}$ and the $HE_{12}$ modes disappears, and the intermodal beatlength between the two modes decreases to a few millimeters or below at any taper diameters (Fig. 1(d)), which is sufficient for the adiabatic transmission of both modes. We examine the criterion on the initial cladding diameter desirable for the adiabatic transmission of the HOMs. The maximum intermodal beatlength between the $EH_{11}$ and the $HE_{12}$ mode over the entire range of possible taper diameter is obtained as a function of the initial cladding diameter, as shown in Fig. 1(e). This result indicates that the fiber cladding should be deeply etched below ~20 μm thickness (roughly twice the core diameter) for the adiabatic transmission of the two modes, where the intermodal beatlength is maintained below ~10 mm during the entire tapering process. We note that in Fig. 1 we also consider the possibility of the coupling between the $HE_{12}$ and the $HE_{31}$ mode at their crossing points. Their coupling is forbidden in an idealized taper having a perfect circular symmetry in the taper cross-section owing to the orthogonality between the azimuthal dependences of the two modes. However, the circular symmetry of the taper cross-section can be easily broken in reality due to the unavoidable imperfection of the tapering process, such as the gas flow, fiber sagging, and flame instability. Then, the two modes can experience intermodal coupling at their crossing points, which would disturb their adiabatic transmission. Nevertheless, in the case of tapering etched fiber, it turns out that the intermodal beatlength between the two modes is also significantly reduced and maintained at a few millimeters or below until the mode-crossing waist diameter (0.7 μm) near their cut-off is reached, as can be seen in Fig. 1(d).

Based on this operating principle, we fabricate silica MASTs employing a two-step process, as described in Fig. 2 (see Methods). We produce MASTs having a waist length of 10 mm and an exponentially shaped transition at both ends of the waist. The length of each taper transition is estimated to be ~60 mm when the waist diameter reaches 0.76 μm. The transmission and the output far-field pattern are recorded for each input HOM during the entire tapering process, as summarized in Figs. 3(b)–3(d) (see Methods). Until the waist diameter reaches ~0.9 μm, the transmission is maintained above 93% for any input HOM. The transmission gets slightly degraded to ~85% when the waist diameter is further reduced, which is attributed mainly to the fact that the taper becomes lifted and curved under the influence of butane-oxygen flow in our current tapering facility, which degrades the thickness uniformity of the taper waist [29,33]. Nevertheless, during the entire tapering process, the far-field pattern of the output beam is not significantly distorted. For the $LP_{02}$ mode (Fig. 3(b)), in particular, the output far-field pattern is almost unchanged until the mode eventually becomes cut-off. On the other hand, the $LP_{21}$ and the $LP_{11}$ modes are not single vector eigenmodes but combinations of {the $EH_{11}$ and $HE_{31}$ modes} and {the $TE_{01}$, $TM_{01}$, and $HE_{21}$ modes}, respectively, leading to the continuous change of the orientation and visibility of intensity lobes in the output field profile during the tapering process.

**High-efficiency fundamental-to-fundamental all-fiber THG in MAST**

The adiabatic transmission of the HOMs through a MAST is useful for several new applications. As an immediate and critical application of the MAST that is capable of robust transmission of the $LP_{02}$ mode, we demonstrate the fundamental-to-fundamental all-fiber phase-matched THG with high conversion efficiencies. It has been theoretically suggested that phase-matching between the fundamental wave in the $HE_{11}$ mode and the third harmonic (TH) in the $HE_{12}$ mode provides the highest THG efficiencies in OMNTs [24]. In relevant previous experiments [17–19], however, the non-adiabatic transmission of the TH signal generated in the $HE_{12}$ mode has seriously deteriorated the THG performance, such as the conversion efficiency and TH beam quality. Figure 4(a) illustrates our experimental scheme of all-fiber THG, where the input pump beam in the $LP_{01}$ mode is eventually converted into the TH in the $LP_{01}$ mode. For efficient

THG, the phase-matching condition should be satisfied, i.e., the effective indices of the fundamental wave and the TH should be the same. We fabricate MASTs with a waist diameter of 0.76 μm (Fig. 2(c)), which is theoretically predicted to yield the phase-matched THG from the $HE_{11}$ mode at 1551 nm wavelength to the $HE_{12}$ mode at 517 nm [24]. The MAST is pumped by optical pulses generated from a master oscillator power amplifier (MOPA) system that utilizes our widely tunable ultra-narrow-linewidth erbium-doped soliton fiber laser [34] as the oscillator. The pump beam is partly converted into the TH in the $HE_{12}$ mode in the MAST, which is then transformed into the desired $LP_{01}$ mode at a home-made all-fiber acoustic-optic mode converter (AOMC) [35] right after the MAST. (see Methods for more details on the THG experimental setup)

Upon pumping each MAST with our wavelength-tunable MOPA system, we routinely observe the intermodally phase-matched THG, as summarized in Figs. 4(b)–4(f). First, the field profile of the $LP_{02}$ mode of the output TH signal is clearly seen when the AOMC is switched off (Fig. 4(b)), thanks to the adiabatic transmission of the $LP_{02}$ mode along the MAST. Furthermore, this TH in the $LP_{02}$ mode can be converted into the $LP_{01}$ mode with high purity (Figs. 4(c)–4(e)), as we turn on the AOMC. We emphasize that this not only further verifies the capability and benefit of adiabatic transmission of the $LP_{02}$ mode of our MAST but also experimentally demonstrates the first-time fundamental-to-fundamental THG in the waveguide platform. To show that the THG is phase-matched, we measure the TH output power while scanning the pump wavelength. The TH power peaks when the pump wavelength is tuned to 1550.80 nm, as shown in Fig. 4(f), which is very close to the theoretically predicted phase-matching pump wavelength of 1550.0 nm. It is noteworthy to mention that the TH signal is also weakly observed over the entire pump wavelength tuning range of the MOPA system (1535–1563 nm). Considering the theoretically anticipated THG bandwidth to be as narrow as 0.3 nm (Fig. 4(g)), we believe that such relatively broadband THG in experiments is due to the non-uniformity of MAST waist diameter. The phase-matching pump wavelength for the THG process is highly sensitive to the waist diameter [36], and our numerical calculation indicates that the variation of waist diameter of only 15 nm gives rise to the shift of phase-matching pump wavelength by an amount of the MOPA tuning range of 28 nm (Fig. 4(h)). Such a small non-uniformity of the waist diameter is attributed to the imperfection of our current tapering facility in the submicron thickness regime that we mentioned previously.

For further verification of the THG, we examine the dependence of the TH power on the pump power and pump polarization, as shown in Fig. 5(a). The TH power is highly dependent on the pump polarization. When we keep adjusting the pump polarization to maximize the TH signal, the measured TH powers exhibit excellent agreement with the well-known cubic dependence on the pump power [24]. The resulting normalized conversion efficiency is determined as $6.6\times10^{-4}$ $W^{-2}$, and the maximum conversion efficiency of $1.5\times10^{-4}$ (71 µW TH output power) is achieved at the pump power of 0.48 W. We also measure the pump power dependence of the TH power when the pump polarization is adjusted to suppress the THG as much as possible. At low pump powers, the TH power can be reduced by a factor of ~1/10, while also displaying a cubic dependence on the pump power. In isotropic media such as fused silica glass, the THG process via $\chi^{(3)}$ nonlinearity becomes most efficient for a linearly polarized pump beam, whereas it is eliminated when the pump beam is circularly polarized [37]. This polarization property should also be exhibited in MASTs, where the TH signal is in the $HE_{12}$ mode that has the same azimuthal dependence of the field distribution as the pump beam in the $HE_{11}$ mode. The nonzero minimum TH signal observed in experiments is attributed to the residual birefringence in the MAST that can arise from slightly broken circular symmetry of the cross-section of the fabricated MAST we mentioned previously, which hinders the preservation of the circular polarization of the pump beam during its propagation along the MAST waist. On the other hand, as the pump power rises further, the THG becomes less polarization-sensitive, with an increase in the minimum TH conversion efficiency. The minimum TH conversion efficiency is locked to about half the maximum value at sufficiently high pump powers, as shown in Fig. 5(b). Such unusual pump polarization dependence might arise from the complicated nonlinear polarization dynamics in the MAST waist [38], which we will investigate in detail in our future study.

Finally, we measure the optical spectrum of the TH signal, together with that of the output pump beam for comparison, as shown in Figs. 5(c)–5(f). At low pump powers (0.13 W), a narrowband TH signal is observed at 516.8 nm and increases as the pump power rises. When the pump power reaches 0.18 W, spectral sidebands are created on both sides at a distance of ~1.6 THz from the main TH peak. Such spectral sidebands also appear in the output pump spectrum with the same spectral distance (~1.6 THz). This indicates that the sidebands are first created in the pump beam via self-phase-modulation (SPM)-induced modulation instability during propagation along the single-mode lead fiber, and then the same sidebands are generated in the TH signal via cross-phase-modulation-induced modulation instability [39] in the MAST rather than the THG of the pump beam sidebands or the SPM-induced modulation instability arising from the main TH peak. We check that such spectral broadening of the pump beam does not appear when the MAST is replaced by intact optical fiber. When the pump power exceeds 0.3 W, both the pump beam and the TH signal experience significant asymmetric spectral broadening (Figs. 5(e) and 5(f)) that arises from nonlinear phase modulation [40].

**DISCUSSION**

We have achieved THG conversion efficiencies which are higher than those reported in most of the previous works. A normalized conversion efficiency of $6.6\times10^{-4}$ $W^{-2}$ and a maximum conversion efficiency of $1.5\times10^{-4}$ at a pump power of 0.48 W are obtained in a MAST waist as short as 10 mm. We expect that the THG efficiency can be further improved by increasing the waist length and the uniformity of the waist diameter. This waist length is currently limited by the taper being lifted up and curved in the tapering process under the influence of the butane-oxygen flow, which deteriorates the thickness uniformity of the taper waist. We observe that wet-etched fibers are more compliant to such gas-flow-induced deformation during the tapering process compared to the case of direct tapering of non-etched conventional optical fibers, which implies that the suppression of the effect of gas flow is essential in the production of high-quality MASTs. We expect that the use of a heat source that does not produce such gas flow, e.g., a focused carbon-dioxide laser beam [41] or an electric micro-heater [42], may resolve the issue and thus enable the fabrication of MASTs with a longer waist and higher transmission.

The unique capability of our MAST to adiabatically transmit several HOMs simultaneously, combined with all-fiber devices that perform the near-unity-efficiency in-fiber mode conversion between the $LP_{01}$ mode and the HOM (at least the $LP_{11}$, $LP_{21}$, and $LP_{02}$ modes), offers a variety of intriguing opportunities in the emerging field of multimode nonlinear optics and quantum optics. Novel multimode nonlinear optical frequency conversion [43] and nonclassical light generation [22,23,25], multimode photonic quantum information processing [44,45], and broadband spatiotemporal dynamics [46] can be investigated with high efficiencies in all-fiber platforms. Furthermore, manipulation of the evanescent field profiles can be facilitated through excitation of HOMs in the MAST waist for tailoring the optical trapping potentials of atoms and nanoparticles [47]. Our MAST can also be used to improve the performance of taper-based photonic devices. For instance, when a saturable absorber is deposited on the MAST waist [48], its operating power can be significantly reduced by transmitting the HOM instead of the $LP_{01}$ mode because of the larger evanescent field of the HOM.

## METHODS

**Fabrication and in-situ characterization of multimode adiabatic submicron tapers (MASTs)**

We fabricate silica MASTs employing a two-step process, as described in Fig. 2(a). A section of unjacketed conventional step-index silica optical fiber, single-mode fiber at telecom wavelength (~1550 nm) with the core diameter of 8.7 μm and the numerical aperture of 0.13, is wet-etched using 6:1 buffered oxide etch solution to reduce the cladding diameter from 125 μm down to below ~20 μm (Fig. 2(b)). Note that the $LP_{11}$, $LP_{21}$, and $LP_{02}$ modes are the core modes at 532 nm wavelength (fiber V parameter: 6.7). The wet-etched fiber section is then tapered to a submicron thickness to produce a final MAST (Fig. 2(c)) via the flame brushing and pulling technique [32] using a butane-oxygen flame as a heat source. To check the adiabatic transmission of each optical mode during the tapering process, we monitor the transmission and far-field pattern simultaneously at the MAST output port, while a laser beam at 532 nm wavelength is coupled into each of the $LP_{02}$, $LP_{21}$, and $LP_{11}$ modes of the MAST input port, as illustrated in Fig. 3(a). To this end, the 532 nm laser beam emitted from a diode-pumped solid-state laser is launched primarily into the $LP_{01}$ mode of the fiber, and the remnant HOMs are removed using a taper-based mode stripper [49] to obtain the pure $LP_{01}$ mode. The $LP_{01}$ mode is then converted into a target HOM using a home-made all-fiber acousto-optic mode converter (AOMC) [35] or a microbend-induced long-period fiber grating [50], whose conversion efficiencies are measured to be as high as ~96% and ~98%, respectively. The transmission and output far-field pattern are recorded for each input HOM using a power meter and a CMOS camera, respectively, during the entire tapering process.

**Observation of fundamental-to-fundamental all-fiber third-harmonic generation (THG) in silica MASTs**

Figure 4(a) illustrates our experimental scheme of fundamental-to-fundamental all-fiber THG. A fabricated MAST is first packaged in an acrylic box for long-term protection from contamination with dust and moisture as well as other environmental perturbations, which is crucial for practical use of MASTs. The MAST is pumped by optical pulses generated from a MOPA system that utilizes our widely tunable ultra-narrow-linewidth dissipative soliton erbium-doped fiber laser [34] as the oscillator. The soliton fiber laser emits an 8-GHz (64-pm)-linewidth, 110-ps-width pulse train at a repetition rate of 2.1 MHz, whose optical wavelength can be tuned over a wide range (1530–1563 nm). The output of the seed laser is amplified using a two-stage erbium-doped fiber amplifier (EDFA) consisting of a home-made pre-amplifier and a power amplifier. A tunable bandpass filter (TBF) having a bandwidth of 0.8 nm is inserted between the two amplifiers to suppress the amplified spontaneous emission noise from the amplifiers, where the center wavelength of the TBF is adjusted according to the laser wavelength. The pulse width of the MOPA output is measured to be 110 ps by using an intensity autocorrelator. We emphasize that the 110 ps pulse width is sufficiently broad compared to the theoretically predicted temporal walk-off of 16 ps between the fundamental wave in the $HE_{11}$ mode and the TH in the $HE_{12}$ mode in the 10-mm-long MAST waist, allowing their temporal overlap to be maintained significantly over the entire MAST waist (see Supplementary Fig. 1). We note that due to the insufficient spectral response of our EDFA at shorter wavelengths, the resultant wavelength tuning range of our MOPA is slightly reduced to 1535–1563 nm. In addition, the maximally attainable pulse peak power for the THG experiments is 1.8 kW, which is mainly limited by the nonlinear spectral broadening at the power amplifier in the two-stage EDFA (see Supplementary Fig. 1). A fiber polarization controller is used after the EDFA to adjust the polarization state of the pump beam for the examination of the pump polarization dependence of the THG process. A 1% fiber tapping coupler is inserted right before the MAST to monitor the pump power. The pump beam is partly converted into the TH in the $HE_{12}$ mode in the MAST, which is then transformed into the desired $LP_{01}$ mode at a home-made all-fiber AOMC [35] right after the MAST. Prior to the THG experiments, we check that a nearly complete (~95%) acousto-optic intermodal conversion is achieved at the AOMC between the $LP_{01}$ and the $LP_{02}$ mode at ~517 nm wavelength when a sinusoidal electric signal of 23 $V_{p-p}$ at 1.404 MHz is applied to the acoustic transducer of the AOMC. The remaining pump beam is rejected right after the fiber output using a short-pass filter. The optical power, far-field pattern, and optical spectrum of the output TH signal are measured with a power meter, a CMOS camera, and a grating-based optical spectrum analyzer, respectively.

# REFERENCES

1. L. Tong, R. R. Gattass, J. B. Ashcom, S. He, J. Lou, M. Shen, I. Maxwell, and E. Mazur, "Subwavelength-diameter silica wires for low-loss optical wave guiding," Nature **426,** 816–819 (2003).

2. G. Brambilla, V. Finazzi, and D. J. Richardson, "Ultra-low-loss optical fiber nanotapers," Opt. Express **12,** 2258–2263 (2004).

3. M. Cai, O. Painter, and K. J. Vahala, "Observation of critical coupling in a fiber taper to a silica-microsphere whispering-gallery mode system," Phys. Rev. Lett. **85,** 74–77 (2000).

4. M. Fujiwara, K. Toubaru, T. Noda, H.-Q. Zhao, and S. Takeuchi, "Highly efficient coupling of photons from nanoemitters into single-mode optical fibers," Nano Lett. **11,** 4362–4365 (2011).

5. R. Yalla, F. L. Kien, M. Morinaga, and K. Hakuta, "Efficient channeling of fluorescence photons from single quantum dots into guided modes of optical nanofiber," Phys. Rev. Lett. **109,** 063602 (2012).

6. E. Vetsch, D. Reitz, G. Sagué, R. Schmidt, S. T. Dawkins, and A. Rauschenbeutel, "Optical interface created by laser-cooled atoms trapped in the evanescent field surrounding an optical nanofiber," Phys. Rev. Lett. **104,** 203603 (2010).

7. B. Gouraud, D. Maxein, A. Nicolas, O. Morin, and J. Laurat, "Demonstration of a memory for tightly guided light in an optical nanofiber," Phys. Rev. Lett. **114,** 180503 (2015).

8. H. L. Sørensen, J.-B. Béguin, K. W. Kluge, I. Iakoupov, A. S. Sørensen, J. H. Müller, E. S. Polzik, and J. Appel, "Coherent backscattering of light off one-dimensional atomic strings," Phys. Rev. Lett. **117,** 133604 (2016).

9. J. Petersen, J. Volz, and A. Rauschenbeutel, "Chiral nanophotonic waveguide interface based on spin-orbit interaction of light," Science **346,** 67–71 (2014).

10. M. Daly, V. G. Truong, and S. Nic Chormaic, "Evanescent field trapping of nanoparticles using nanostructured ultrathin optical fibers," Opt. Express **24,** 14470–14482 (2016).

11. W. Jin, H. L. Ho, Y. C. Cao, J. Ju, and L. F. Qi, "Gas detection with micro- and nano-engineered optical fibers," Opt. Fiber Technol. **19,** 741–759 (2013).

12. Z. Ding, K. Sun, K. Liu, J. Jiang, D. Yang, Z. Yu, J. Li, and T. Liu, "Distributed refractive

index sensing based on tapered fibers in optical frequency domain reflectometry," Opt. Express **26,** 13042–13054 (2018).

13. S. G. Leon-Saval, T. A. Birks, W. J. Wadsworth, P. St. J. Russell, and M. W. Mason, "Supercontinuum generation in submicron fibre waveguides," Opt. Express **12,** 2864–2869 (2004).

14. M. S. Kang, A. Brenn, G. S. Wiederhecker, and P. St. J. Russell, "Optical excitation and characterization of gigahertz acoustic resonances in optical fiber tapers," Appl. Phys. Lett. **93,** 131110 (2008).

15. L. Cui, X. Li, C. Guo, Y. H. Li, Z. Y. Xu, L. J. Wang, and W. Fang, "Generation of correlated photon pairs in micro/nano-fibers," Opt. Lett. **38,** 5063–5066 (2013).

16. J. Kim, Y. S. Ihn, Y. Kim, and H. Shin, "Photon-pair source working in a silicon-based detector wavelength range using tapered micro/nanofibers," Opt. Lett. **44,** 447–450 (2019).

17. V. Grubsky and J. Feinberg, "Phase-matched third-harmonic UV generation using low-order modes in a glass micro-fiber," Opt. Commun. **274,** 447–450 (2007).

18. T. Lee, Y. Jung, C. A. Codemard, M. Ding, N. G. R. Broderick, and G. Brambilla, "Broadband third harmonic generation in tapered silica fibres," Opt. Express **20,** 8503–8511 (2012).

19. A. Coillet and P. Grelu, "Third-harmonic generation in optical microfibers: From silica experiments to highly nonlinear glass prospects," Opt. Commun. **285,** 3493–3497 (2012).

20. M. I. M. Abdul Khudus, T. Lee, F. D. Lucia, C. Corbari, P. Sazio, P. Horak, and G. Brambilla, "All-fiber fourth and fifth harmonic generation from a single source," Opt. Express **24,** 21777–21793 (2016).

21. Y. Wang, T. Lee, F. De Lucia, M. I. M. Abdul Khudus, P. J. A. Sazio, M. Beresna, and G. Brambilla, "All-fiber sixth-harmonic generation of deep UV," Opt. Lett. **42,** 4671–4674 (2017).

22. M. Corona, K. Garay-Palmett, and A. B. U'Ren, "Third-order spontaneous parametric down-conversion in thin optical fibers as a photon-triplet source," Phys. Rev. A **84,** 033823 (2011).

23. A. Dot, A. Borne, B. Boulanger, K. Bencheikh, and J. A. Levenson, "Quantum theory analysis of triple photons generated by a $\chi^{(3)}$ process," Phys. Rev. A **85,** 023809 (2012).

24. V. Grubsky and A. Savchenko, "Glass micro-fibers for efficient third harmonic generation," Opt. Express **13,** 6798–6806 (2005).

25. S. Signorini, M. Mancinelli, M. Borghi, M. Bernard, M. Ghulinyan, G. Pucker, and L. Pavesi, "Intermodal four-wave mixing in silicon waveguides," Photon. Res. **6,** 805–814 (2018).

26. V. Mishra, S. P. Singh, R. Haldar, P. Mondal, and S. K. Varshney, "Intermodal nonlinear effects mediated optical event horizon in short-length multimode fiber," Phys. Rev. A **96,** 013807 (2017).

27. L. Rishøj, B. Tai, P. Kristensen, and S. Ramachandran, "Soliton self-mode conversion: revisiting Raman scattering of ultrashort pulses," Optica **6,** 304–308 (2019).

28. S. Ravets, J. E. Hoffman, L. A. Orozco, S. L. Rolston, G. Beadie, and F. K. Fatemi, "A low-loss photonic silica nanofiber for higher-order modes," Opt. Express **21,** 18325–18335 (2013).

29. J. M. Ward, A. Maimaiti, V. H. Le, and S. Nic Chormaic, "Optical micro- and nanofiber pulling rig," Rev. Sci. Instrum. **85,** 111501 (2014).

30. K. Harrington, S. Yerolatsitis, D. V. Ras, D. M. Haynes, and T. A. Birks, "Endlessly adiabatic fiber with a logarithmic refractive index distribution," Optica **4,** 1526–1533 (2017).

31. Y. Jung, K. Harrington, S. Yerolatsitis, D. J. Richardson, and T. A. Birks, "Adiabatic higher-order mode microfibers based on a logarithmic index profile," Opt. Express **28,** 19126–19132 (2020).

32. T. A. Birks and Y. W. Li, "The shape of fiber tapers," J. Lightwave Technol. **10,** 432–438 (1992).

33. Y. Xu, W. Fang, and L. Tong, "Real-time control of micro/nanofiber waist diameter with ultrahigh accuracy and precision," Opt. Express **25,** 10434–10440 (2017).

34. C. K. Ha, K. S. Lee, D. Kwon, and M. S. Kang, "Widely tunable ultra-narrow-linewidth dissipative soliton generation at telecom band," Photon. Res. **8,** 1100–1109 (2020).

35. J. Zhao, R. Miao, and X. Liu, "Nonresonant cascaded acousto-optic mode coupling," Opt. Lett. **31,** 2909–2911 (2006).

36. M. I. M. Abdul Khudus, T. Lee, P. Horak, and G. Brambilla, "Effect of intrinsic surface roughness on the efficiency of intermodal phase matching in silica optical nanofibers," Opt. Lett. **40,** 1318–1321 (2015).

37. P. P. Bey, J. H. Giuliani, and H. Rabin, "Linear and circular polarized laser radiation in optical third harmonic generation," Phys. Lett. **26A,** 128–129 (1968).

38. D. Hartley, M. A. Lohe, T. M. Monro, and S. Afshar V., "Cross mode and polarization mixing in third and one-third harmonic generation in multimode waveguides," J. Opt. Soc. Am. B **32,** 379–387 (2015).

39. Y. B. Band, "Phase-modulation effects on the spectrum of third-harmonic generation," Phys. Rev. A **42,** 5530–5536 (1990).

40. A. N. Naumov and A. M. Zheltikov, "Asymmetric spectral broadening and temporal evolution of cross-phase-modulated third-harmonic pulses," Opt. Express **10,** 122–127 (2002).

41. H. Yokota, E. Sugai, and Y. Sasaki, "Optical irradiation method for fiber coupler fabrications," Opt. Rev. **4,** 104–107 (1997).

42. L. Ding, C. Belacel, S. Ducci, G. Leo, and I. Favero, "Ultralow loss single-mode silica tapers manufactured by a microheater," Appl. Opt. **49,** 2441–2445 (2010).

43. J. Demas, L. Rishøj, X. Liu, G. Prabhakar, and S. Ramachandran, "Intermodal group-velocity engineering for broadband nonlinear optics," Photon. Res. **7,** 1–7 (2019).

44. L.-T. Feng, M. Zhang, Z.-Y. Zhou, M. Li, X. Xiong, L. Yu, B.-S. Shi, G.-P. Guo, D.-X. Dai, X.-F. Ren, and G.-C. Guo, "On-chip coherent conversion of photonic quantum entanglement between different degrees of freedom," Nat. Commun. **7,** 11985 (2016).

45. L.-T. Feng, M. Zhang, X. Xiong, Y. Chen, H. Wu, M. Li, G.-P. Guo, G.-C. Guo, D.-X. Dai, X.-F. Ren, "On-chip transverse-mode entangled photon pair source," NPJ Quantum Inf. **5,** 2 (2019).

46. L. G. Wright, D. N. Christodoulides, and F. W. Wise, "Spatiotemporal mode-locking in multimode fiber lasers," Science **358,** 94–97 (2017).

47. A. Maimaiti, V. G. Truong, M. Sergides, I. Gusachenko, and S. Nic Chormaic, "Higher order microfiber modes for dielectric particle trapping and propulsion," Sci. Rep. **5,** 9077 (2015).

48. A. Martinez, M. A. Araimi, A. Dmitriev, P. Lutsyk, S. Li, C. Mou, A. Rozhin, M. Sumetsky, and S. Turitsyn, "Low-loss saturable absorbers based on tapered fibers embedded in carbon nanotube/polymer composites," APL Photon. **2,** 126103 (2017).

49. D. Đonlagic, "In-line higher order mode filters based on long highly uniform fiber tapers," J. Lightwave Technol. **24,** 3532–3539 (2006).

50. J. N. Blake, B. Y. Kim, and H. J. Shaw, "Fiber-optic modal coupler using periodic microbending," Opt. Lett. **11,** 177–179 (1986).

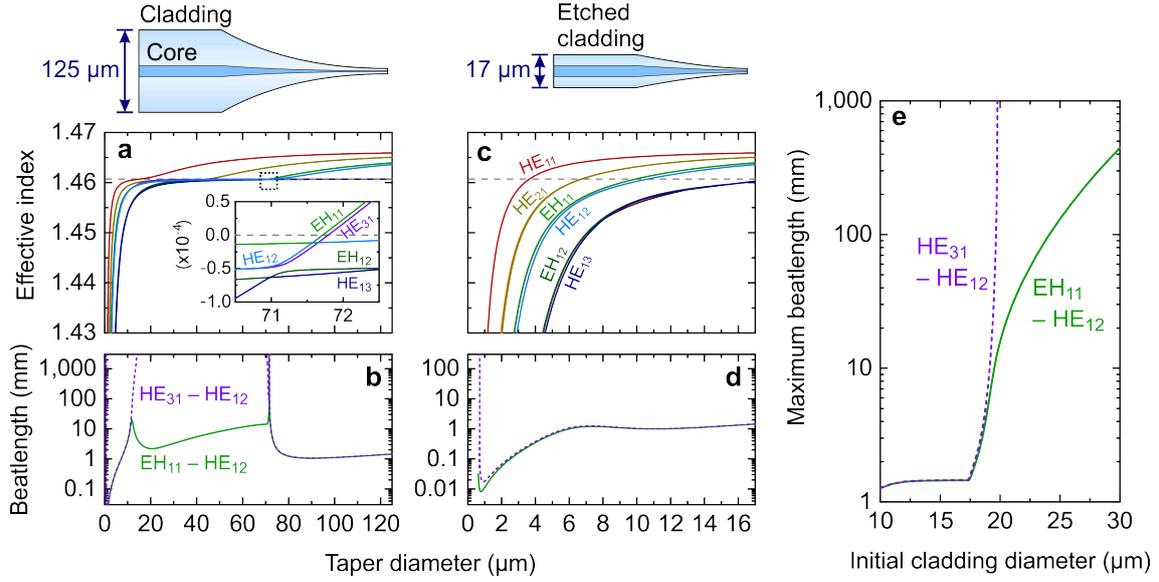

**FIG. 1. Operating principle of multimode adiabatic submicron taper.** (a) Calculated effective indices of several modes over a range of taper diameter in the case when a 125-μm-thick step-index silica fiber (core diameter: 8.7 μm, numerical aperture: 0.13) is directly tapered. The inset is the zoomed-in plot around the black dotted box, where the vertical axis represents the 'relative' effective index from the refractive index of silica cladding (grey horizontal dashed lines in (a) and (c)). (b) Intermodal beatlength between the $EH_{11}$ and the $HE_{12}$ mode (green solid curve) and that between the $HE_{31}$ and the $HE_{12}$ mode (violet dotted curve) obtained from (a). (c) Calculated effective indices in the scenario when a 17-μm-thick cladding-etched fiber is tapered. (d) Intermodal beatlength obtained from (c). (e) Maximum intermodal beatlength during the tapering process as a function of the initial diameter of etched cladding. All calculations are performed at 532 nm wavelength.

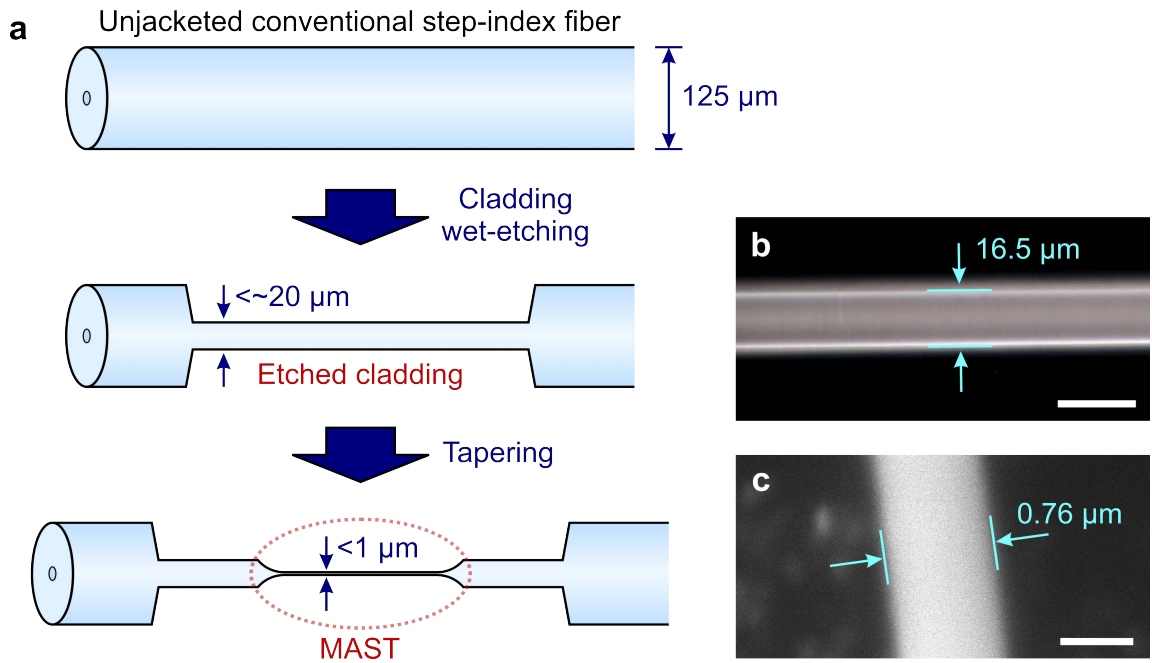

**FIG. 2. Fabrication of multimode adiabatic submicron tapers (MASTs).** (a) Fabrication process of a silica MAST. A section of unjacketed conventional step-index silica optical fiber is etched to a thickness below ~20 μm. The etched cladding is then tapered to a submicron thickness using the flame brushing and pulling technique [32]. (b) Optical micrograph of a 16.5-μm-thick cladding-etched fiber. The white scale bar corresponds to 20 μm. (c) Scanning electron micrograph of a 0.76-μm-thick silica MAST. The white scale bar corresponds to 500 nm.

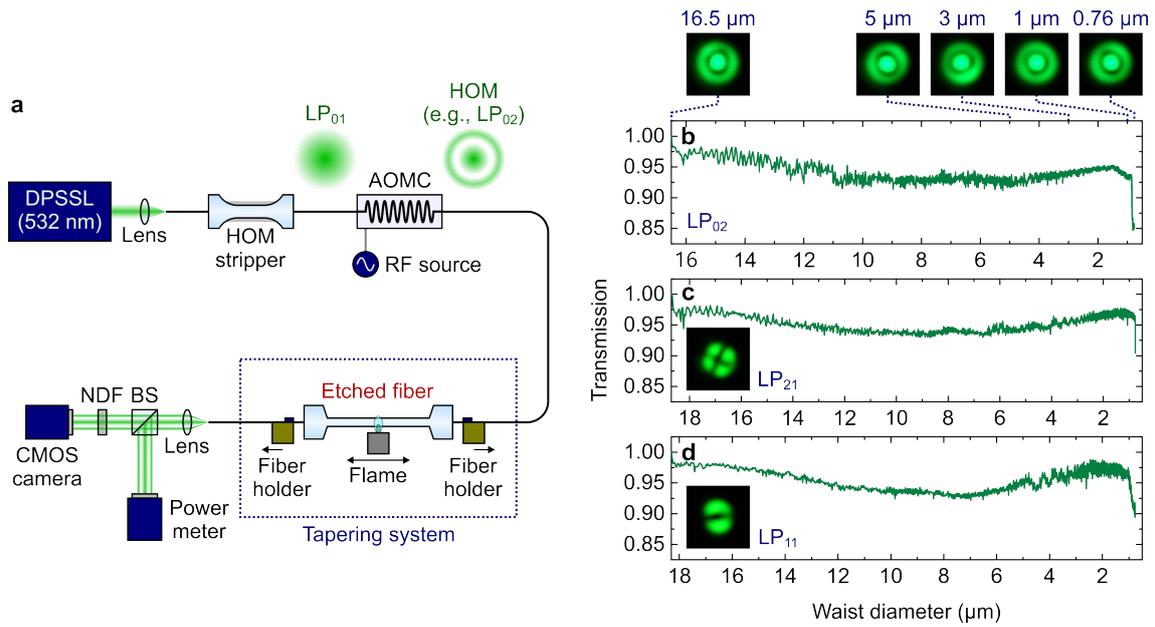

**FIG. 3. In-situ optical characterization of multimode adiabatic submicron tapers (MASTs) during the tapering process.** (a) Schematic diagram of the experimental setup for the fabrication and in-situ transmission measurement of a silica MAST. A higher-order mode (HOM) at 532 nm wavelength is generated through in-fiber acousto-optic mode conversion from the input $LP_{01}$ mode and then launched into the tapering system. The transmission and output far-field pattern are monitored with a power meter and a CMOS camera, respectively, during the tapering process. DPSSL, diode-pumped solid-state laser; AOMC, acousto-optic mode converter; BS, 50/50 beam splitter; NDF, neutral density filter. (b–d) Transmission and output far-field pattern that are recorded during the entire tapering process, when the $LP_{02}$ mode (b), the $LP_{21}$ mode (c), or the $LP_{11}$ mode (d) is launched into the input port. In (c) and (d), the CMOS images are the far-field patterns before the tapering process starts.

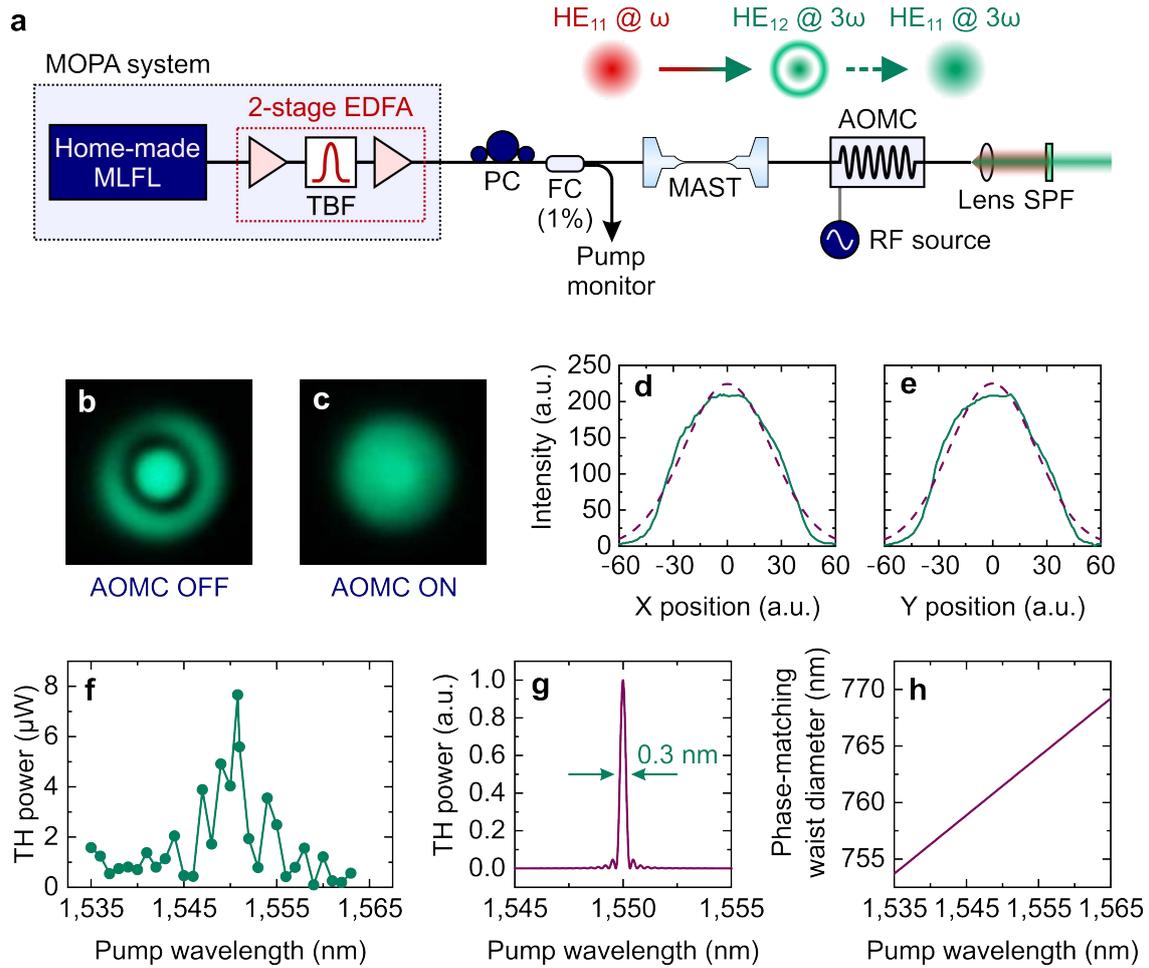

**FIG. 4. Fundamental-to-fundamental all-fiber phase-matched third-harmonic generation in multimode adiabatic submicron tapers (MASTs).** (a) Schematic diagram of the experimental setup. Red and turquoise in the beam path indicate the ~1550 nm pump beam and the third harmonic (TH), respectively. MLFL, mode-locked fiber laser; EDFA, erbium-doped fiber amplifier; TBF, tunable bandpass filter; PC, polarization controller; FC, fiber coupler; AOMC, acousto-optic mode converter; SPF, short-pass filter. (b,c) Measured far-field patterns of the output TH when the AOMC is turned off (b) and switched on (c). (d,e) Gaussian fits (purple dashed curves) to the experimentally obtained intensity profiles (turquoise solid curves) from (c). (f) TH output powers measured at an average pump power of 0.14 W over a range of pump wavelength. (g) Theoretically predicted TH power as a function of pump wavelength with an assumption of uniform waist diameter (0.76 μm). (h) Calculated phase-matching waist diameter as a function of pump wavelength.

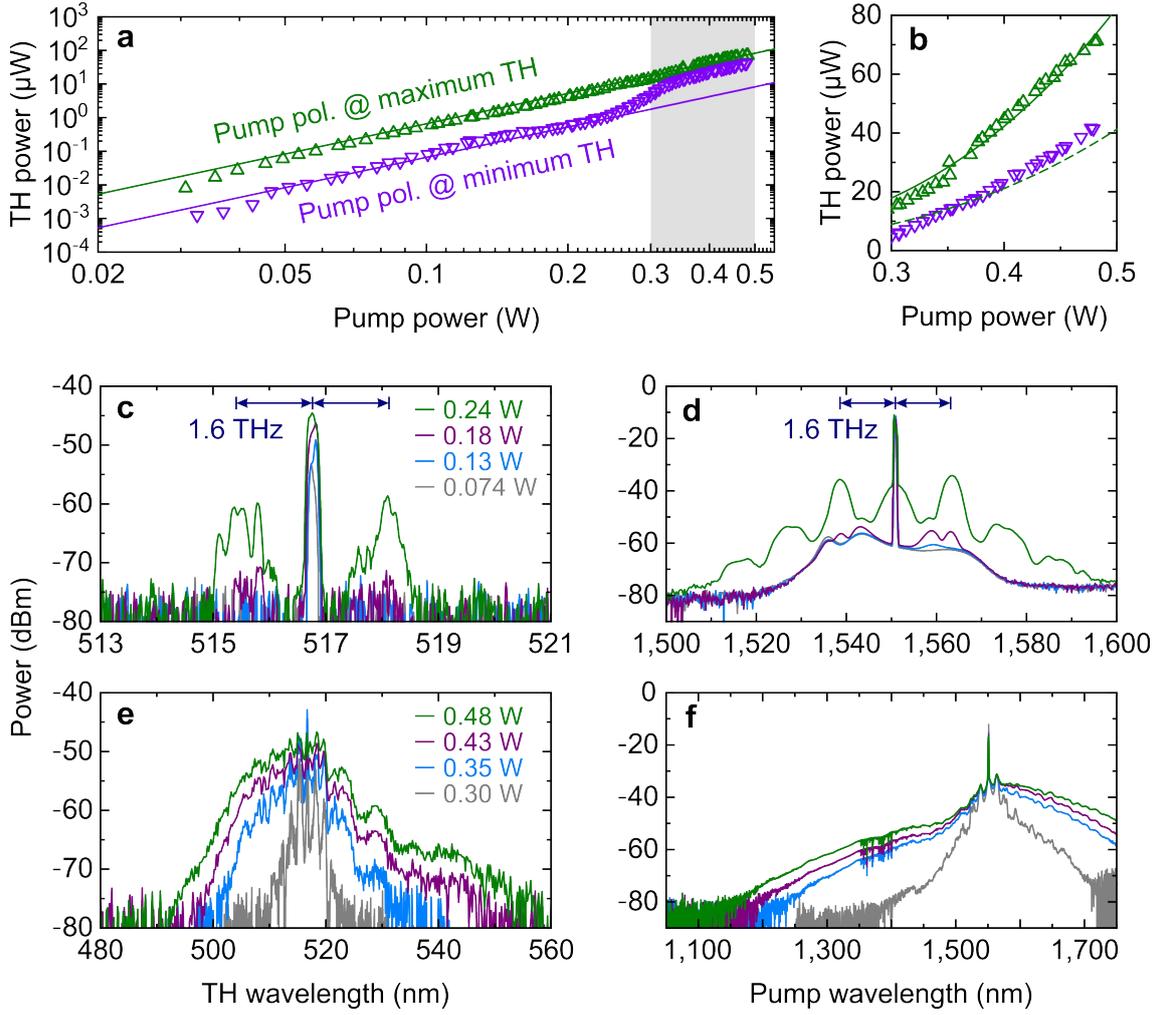

**FIG. 5. Pump power dependence of third-harmonic (TH) generation.** (a) Measured TH powers over a range of average pump power, while the input pump polarization is adjusted to yield the maximum (green triangles) or minimum (violet triangles) TH power for each pump power. The green and violet lines are the cubic fits to the measured maximum and minimum TH powers, respectively. (b) Zoomed-in plot around the grey region in (a). The green solid curve is the same as the cubit fit (green line) in (a), and the green dashed curve stands for half of the cubit fit. (c,d) Optical spectra of the TH (c) and output pump beam (d) at relatively low pump powers. (e,f) Optical spectra of the TH (e) and output pump beam (f) at relatively high pump powers. Each color in (d) and (f) corresponds to the average pump power displayed in (c) and (e), respectively.

# Supplementary Note: Characteristics of the master oscillator power amplifier system for third-harmonic generation experiments

In our third-harmonic generation (THG) experiments, the multimode adiabatic submicron taper (MAST) is pumped by optical pulses generated from a master oscillator power amplifier (MOPA) system that employs our widely tunable ultra-narrow-linewidth dissipative soliton erbium-doped fiber laser [1] as the oscillator (see Fig. 4(a) in the main text). The soliton fiber laser emits an 8-GHz (64-pm)-linewidth, 110-ps pulse train at 2.1 MHz repetition rate. The pulse width of the MOPA output is also measured to be 110 ps by using an intensity autocorrelator (Supplementary Fig. 1(a)). We emphasize that the 110 ps pulse width is sufficiently broad compared to the theoretically predicted temporal walk-off of 16 ps between the fundamental wave in the $LP_{01}$ mode and the third harmonic (TH) in the $LP_{02}$ mode in the 10-mm-long MAST waist (Supplementary Fig. 1(b)), allowing the significant temporal overlap between the pump pulse and the TH signal to be maintained over the entire MAST waist. On the other hand, the maximum attainable pulse peak power for the THG experiment is 1.8 kW, which is mainly limited by the self-phase-modulation-induced spectral broadening at the power amplifier in the two-stage erbium-doped fiber amplifier. The resultant spectral width of the MOPA output increases up to 0.44 nm at such high peak powers (Supplementary Fig. 1(c)), which is comparable to the theoretically predicted bandwidth of 0.3 nm for the THG process in the 10-mm-long MAST waist (Supplementary Fig. 1(d)).

# References

1. C. K. Ha, K. S. Lee, D. Kwon, and M. S. Kang, "Widely tunable ultra-narrow-linewidth dissipative soliton generation at the telecom band," Photon. Res. **8,** 1100–1109 (2020).

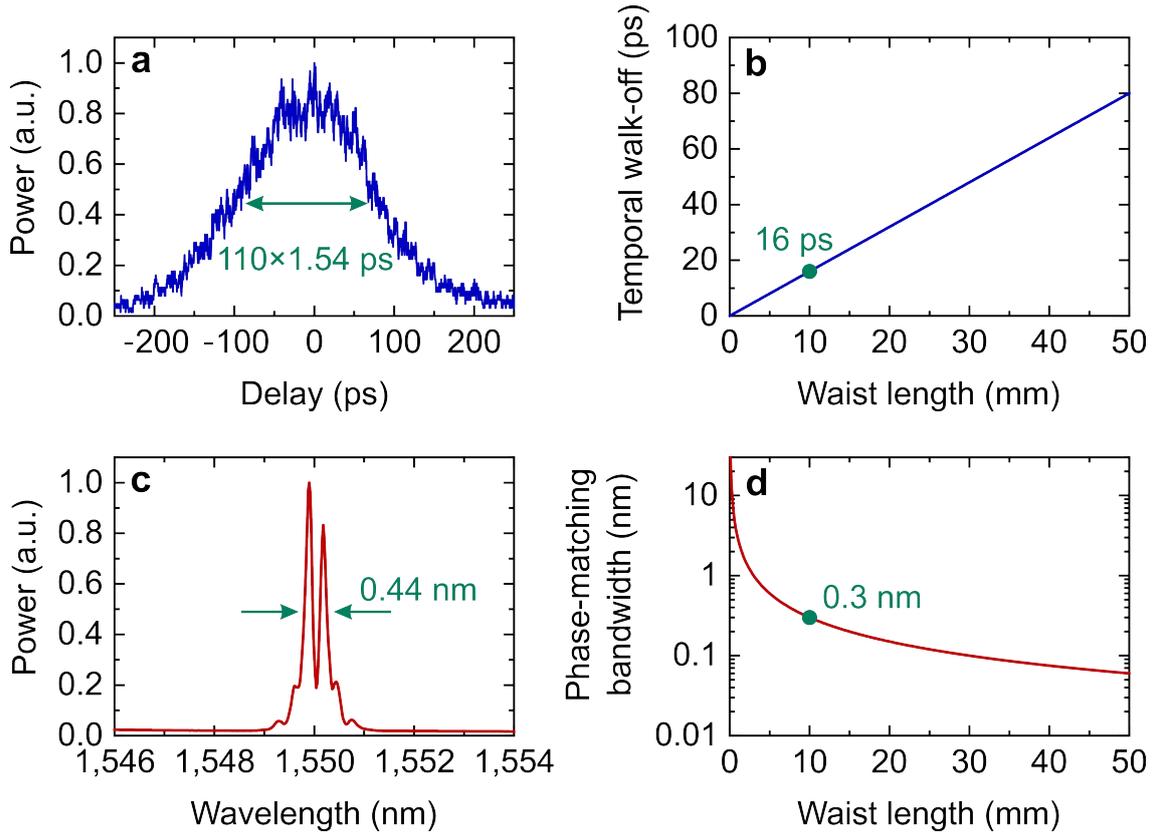

**Supplementary Figure 1. Characteristics of the pump pulses generated from the master oscillator power amplifier (MOPA) system at the maximum pulse peak power (1.8 kW) for third-harmonic generation (THG) experiments.** (a) Intensity autocorrelation trace. (b) Theoretically predicted temporal walk-off between the 1550 nm pump pulse and the 517 nm third harmonic (TH) in a multimode adiabatic submicron taper (MAST) of the waist thickness of 0.76 μm, as a function of the waist length. (c) Optical spectrum measured with a grating-based optical spectrum analyzer. (d) Theoretically predicted phase-matching bandwidth of THG as a function of the waist length. In (b) and (d), the green solid circles correspond to the 10-mm-long waist in our THG experiments. In (a) and (b), the 110 ps pulse width is sufficiently broader than the theoretically predicted temporal walk-off of 16 ps in our THG experiments, allowing the temporal overlap between the pump pulse and the TH signal to be maintained significantly over the entire MAST waist. In (c) and (d), the 3-dB spectral width of the MOPA output increases up to 0.44 nm via self-phase modulation (SPM) at such high pulse peak powers, which is comparable to the 3-dB phase-matching bandwidth of 0.3 nm for the THG. The nonlinear spectral broadening is the main limiting factor of the attainable pulse peak power for our THG experiments.